\documentclass[%
 reprint,
 amsmath,amssymb,
 aps,
]{revtex4-2}

\usepackage{graphicx}
\usepackage{dcolumn}
\usepackage{bm}
\usepackage{balance}

\begin{document}

\preprint{APS/123-QED}

\title{Collective filament wrapping and nested spiral formation \\ in active polydisperse systems}

\author{Caterina Landi}%
\affiliation{%
 Department of Structure of Matter, Thermal Physics, and Electronics, Complutense University of Madrid, Madrid, 28040, Spain.
}%

\author{Giulia Janzen}%
\affiliation{%
 Department of Theoretical Physics, Complutense University of Madrid, Madrid, 28040, Spain 
}%

\author{Francesco Sciortino}
\author{John Russo}
\affiliation{
 Department of Physics, Sapienza Universit\'a di Roma, 00185 Rome, Italy.
}%

\author{Chantal Valeriani}%
\affiliation{%
 Department of Structure of Matter, Thermal Physics, and Electronics, Complutense University of Madrid, Madrid, 28040, Spain.
}%

\author{Daniel A. Matoz-Fernandez }
\email{Correspondence should be addressed to dmatoz@ucm.es}
\affiliation{%
 Department of Theoretical Physics, Complutense University of Madrid, Madrid, 28040, Spain 
}%

\date{\today}

\begin{abstract}
We investigate a two-dimensional polydisperse suspension of self-propelled semiflexible filaments and reveal a collective wrapping mechanism that is absent in monodisperse systems. At intermediate activity levels, long filaments coil around shorter ones, forming nested spiral structures stabilized by filament length disparity. These assemblies generalize the single-filament spiraling seen in active systems into cooperative, multi-filament configurations. As activity increases, the nested spirals undergo structural transitions: medium-length filaments unwind, longer filaments encapsulate shorter ones, and eventually all spiral structures dissolve. This reorganization is reflected in the dynamics, where van Hove distributions uncover coexisting confined and motile filament populations. Our findings identify filament length as a key control parameter for nonequilibrium self-assembly and establish inter-filament wrapping as a minimal mechanism for hierarchical organization in active matter. This mechanism provides a simple model for the cooperative confinement and structural hierarchy observed in both biological and synthetic active systems.
\end{abstract}

\maketitle

\section{Introduction}

Filamentous active systems are found across biology and synthetic materials, exhibiting rich dynamics driven by their elongated shape and self-propulsion~\cite{metareview}. In biological contexts, they include cytoskeletal filaments such as microtubules and actin~\cite{alfredo_new, ganguly2012cytoplasmic}, as well as larger organisms like elongated bacteria~\cite{Yaman2019, serena2019, auer_bacterial_2019} and worms~\cite{deblais2020phase, deblais2020rheology, nguyen2021emergent}. Synthetic analogues range from active chains of metal-dielectric Janus particles under electric fields~\cite{yan_2016}, to vibrated granular rods with polymer-like behavior~\cite{wen2012polymerlike, soh2019self}, and robotic swarms designed for collective motion~\cite{marvi2014sidewinding, ozkan2021collective}.

Among the many collective behaviors exhibited by active filaments, the emergence of swirls, vortices, and spiral patterns stands out as particularly intriguing. In cytoskeletal systems, such patterns have been observed across a range of experimental setups~\cite{bourdieu_spiral_1995, Schaller2010, sumino_2012, sciortino_2021}, often serving functional roles. For instance, cortical microtubules in plant cells organize into swirling arrays that guide anisotropic cell wall growth~\cite{cortical_microtubule}. Similar spiral structures have also been reported in dense bacterial suspensions, where longer filamentous bacteria are surrounded by shorter ones~\cite{lin_2014}, highlighting the role of filament flexibility in spiral pattern formation.

Numerical simulations in two dimensions have shown that individual self-propelled, flexible filaments can spontaneously form stable spiral conformations~\cite{IseleHolder2015}. In monodisperse two-dimensional systems, increasing activity induces a transition from open-chain swimming to frozen spiral states~\cite{duman2018collective}, which subsequently break apart at even higher activity levels~\cite{janzen2024density}. 
Similar stable spiral conformations have been observed in dense filament layers~\cite{Prathyusha2018}, in filaments driven by explicit motor proteins~\cite{Shee2021}, and in chains of chiral active Brownian particles~\cite{anand_2025}. In three dimensions, active filaments exhibit activity-dependent behaviors distinct from spiral conformations, such as coil-to-globule-like transitions~\cite{Bianco2108} and reentrant swelling~\cite{Li2023}.

Despite the extensive phenomenology uncovered in active filament systems~\cite{winkler_physics_2020}, most theoretical and simulation studies to date have relied on monodisperse models, where all filaments share the same length. In contrast, real-world systems are inherently polydisperse. In biological contexts, filament length varies due to dynamic processes such as polymerization, depolymerization, severing, and annealing~\cite{gopinathan_2007, pavlov_2007, humphrey_2002},
while in engineered systems, both synthetic polymers~\cite{lietor_2010} and self-morphic active materials~\cite{kumar_2024} frequently display length heterogeneity.

Polydispersity introduces new forms of structural and dynamical heterogeneity, which can fundamentally alter collective behavior~\cite{Li2025Polydispersity, de_filippo2023}. 
Yet its role in active filament systems remains largely unexplored. By explicitly incorporating length variability into our models, we aim to capture more realistic dynamics and reveal new mechanisms of self-organization. A recent study~\cite{landi2024self} demonstrated that active Brownian particles assembling into polydisperse chains exhibit distinct phase behavior—ranging from crystalline clusters to motility-induced spirals—depending on activity and temperature. However, because these chains form and break dynamically, their structural characterization remains challenging.

Motivated by the gap in understanding spiral formation under polydispersity, we study a system of tangentially self-propelled filaments with fixed bond lengths and tunable length distributions, as detailed in Section~\ref{sec:model}. Our goal is to investigate how filament length heterogeneity influences collective dynamics—particularly the formation, stability, and evolution of spiral structures. As shown in Section~\ref{sec:turning_number}, polydispersity does not suppress spiral formation or the characteristic reentrant behavior observed in monodisperse systems. Instead, it shifts the transition thresholds for different filament lengths, enabling spiral configurations to persist at higher activity levels. This extended stability arises from a cooperative trapping mechanism, where long filaments dynamically wrap around and confine shorter ones. In Section~\ref{nested}, we demonstrate that increasing activity drives a gradual transition from multi-filament to two-filament nested spirals, reflecting the fact that filaments of different lengths unwind at different Péclet numbers. This structural reorganization is mirrored in the dynamics: van Hove displacement distributions reveal coexisting populations of confined and motile filaments, depending on their length and degree of entanglement.

Together, these findings show that polydispersity promotes a hierarchical mode of self-assembly, where structural diversity enables cooperative confinement and the stabilization of collective wrapping across activity regimes.

\section{Model} 
\label{sec:model}

We investigate a two-dimensional polydisperse system of self-propelled filaments to study how length heterogeneity affects their structural organization and dynamical behavior. Each filament is modeled as a linear chain of $N_b$ active beads. The number of beads per filament remains fixed; filaments neither break nor recombine. The system is simulated in the dry limit, where long-range hydrodynamic interactions are neglected. The surrounding medium exerts only local friction and thermal noise. The motion of each bead follows the underdamped Langevin equation~\cite{IseleHolder2015, Prathyusha2018}:
\begin{equation}
m_i \, \ddot{\mathbf{r}_i}=-\gamma \, \dot{\mathbf{r}_i}+\mathbf{R}_i(t)+\sum_{j\neq i } \mathbf{f}_{ij}+\mathbf{f}^a_i
\end{equation}
where $m_i$ and $\mathbf{r}_i$ represent the mass and the position in space of the bead $i$, dots indicate time derivatives, and $\gamma$ is the damping coefficient. The term $\mathbf{R}_i(t)$ represents a delta-correlated random force, characterized by a zero mean and a variance of $4 \gamma k_B T \delta_{ij} \delta(t-t^\prime)$, where $k_B$ is the Boltzmann constant, and $T$ is the temperature.
The interaction force is derived from the total potential $\phi(r_{ij})$, such that $\mathbf{f}_{ij} = -\nabla_i \phi(r_{ij})
$, where $r_{ij} = |\mathbf{r}_i - \mathbf{r}_j|$. The total potential is decomposed as $\phi(r_{ij}) = \phi_{\mathrm{B}}(r_{ij}) + \phi_{\mathrm{NB}}(r_{ij})$, with $\phi_{\mathrm{B}}$ accounting for bonded interactions and $\phi_{\mathrm{NB}}$ for short-range non-bonded steric repulsion. Bonded interactions include a stretching contribution modeled via the Tether potential~\cite{Noguchi2005} and a bending contribution given by a harmonic angle potential~\cite{Prathyusha2018}, such that $\phi_{\mathrm{B}} = \phi_{\mathrm{bond}} + \phi_{\mathrm{bend}}$. Note that, in the regime of small bond-length fluctuations studied here, the choice between the Tether and FENE potentials has negligible impact on system behavior~\cite{janzen2024density}. Non-bonded interactions $\phi_{\mathrm{NB}}$ are modeled via the Weeks–Chandler–Andersen (WCA) potential~\cite{weeks1971role}, ensuring purely repulsive excluded-volume effects.
The self-propulsion force acting on bead $i$ is defined as $\mathbf{f}_i^a = f_a \, \mathbf{t}_{i-1,i+1}$, where $f_a$ is the propulsion strength, and $\mathbf{t}_{i-1,i+1} = \frac{\mathbf{r}_{i+1} - \mathbf{r}_{i-1}}{|\mathbf{r}_{i+1} - \mathbf{r}_{i-1}|}$ is a unit vector approximating the local backbone tangent~\cite{jiang2014motion, Bianco2108, Foglino2019}. End beads ($i=1$ and $i=N_b$) do not experience propulsion, i.e., $f_a = 0$ at the filament ends. Note that choosing the push–pull active polymer model \cite{IseleHolder2015, Prathyusha2018} would yield similar results, provided the mean active force is appropriately normalized \cite{janzen2024active}. 

Finally, the system contains $N = 5 \times 10^4$ particles. We consider filaments of various lengths, with the number of beads per filament given by $N_b \in \{3,\ 11,\ 20,\ 28,\ 37,\ 45,\ 54,\ 62,\ 71,\ 80\}$. The system contains an equal number of filaments for each length. A single filament may slightly deviate from this set to ensure that the total number of particles exactly matches $N$. 
The length of a filament is approximated as $L \approx b (N_b - 1)\sigma$, with $b = 0.86$ the average bond length. The stiffness of each filament is characterized by the dimensionless persistence ratio $\xi_p / L = 2b\kappa / (Lk_B T)$, where $\kappa$ is the bending rigidity. We fix $\kappa$ such that $\xi_p / L = 1.3$ for all filaments.
The strength of active forcing relative to thermal noise is quantified by the Péclet number $Pe = \frac{3 f_a \sigma}{k_B T}$. Simulations are performed at a fixed packing fraction $\rho = 0.3$, and $Pe$ is varied to explore different activity regimes. 

Simulation units are set by choosing the bead diameter $\sigma = 1.0$, energy scale $\epsilon = 1.0$, thermal energy $k_B T = 0.1$, and damping coefficient $\gamma = 1.0$.
To ensure that the system is in steady-state, simulations are run for $8 \times 10^4$ steps with a time step of $\delta t=10^{-3}$. For $Pe \geq 16.0$, a smaller time step  of $\delta t=5\times10^{-4}$ is used to maintain numerical stability. Static properties are then averaged over $100$ configurations sampled every $500$ steps. All simulations are carried out using the GPU-accelerated \textsc{SAMoS} molecular dynamics package~\cite{SAMoS2024}, with time integration performed using the BAOAB scheme for Langevin dynamics~\cite{leimkuhler2015molecular}.
\section{Results and discussion}
\label{sec:results}
\subsection{Structural characterization: turning number per filament length}
\label{sec:turning_number}

To characterize the system's structural behavior, we compute the turning number~\cite{krantz1999handbook,Shee2021} for each filament and then average these values over filaments of the same length. The turning number is defined as
\begin{equation}
    \psi = \frac{1}{2 \pi} \sum_{j=1}^{N_b-1} (\theta_{j+1}-\theta_j),
\label{eq:turning_number}
\end{equation}
where $\theta_{j+1} - \theta_j$ represents the angular variation between two consecutive beads. This quantity measures the number of turns that a filament makes between its two ends: for an elongated filament, $|\psi| = 0$ (no turn); for a circular filament, $|\psi| = 1$ (one turn); and higher values of $|\psi|$ correspond to filaments in spiral configurations (more than one turn). 

\begin{figure}[h!]
    \centering
    \includegraphics[width=\columnwidth]{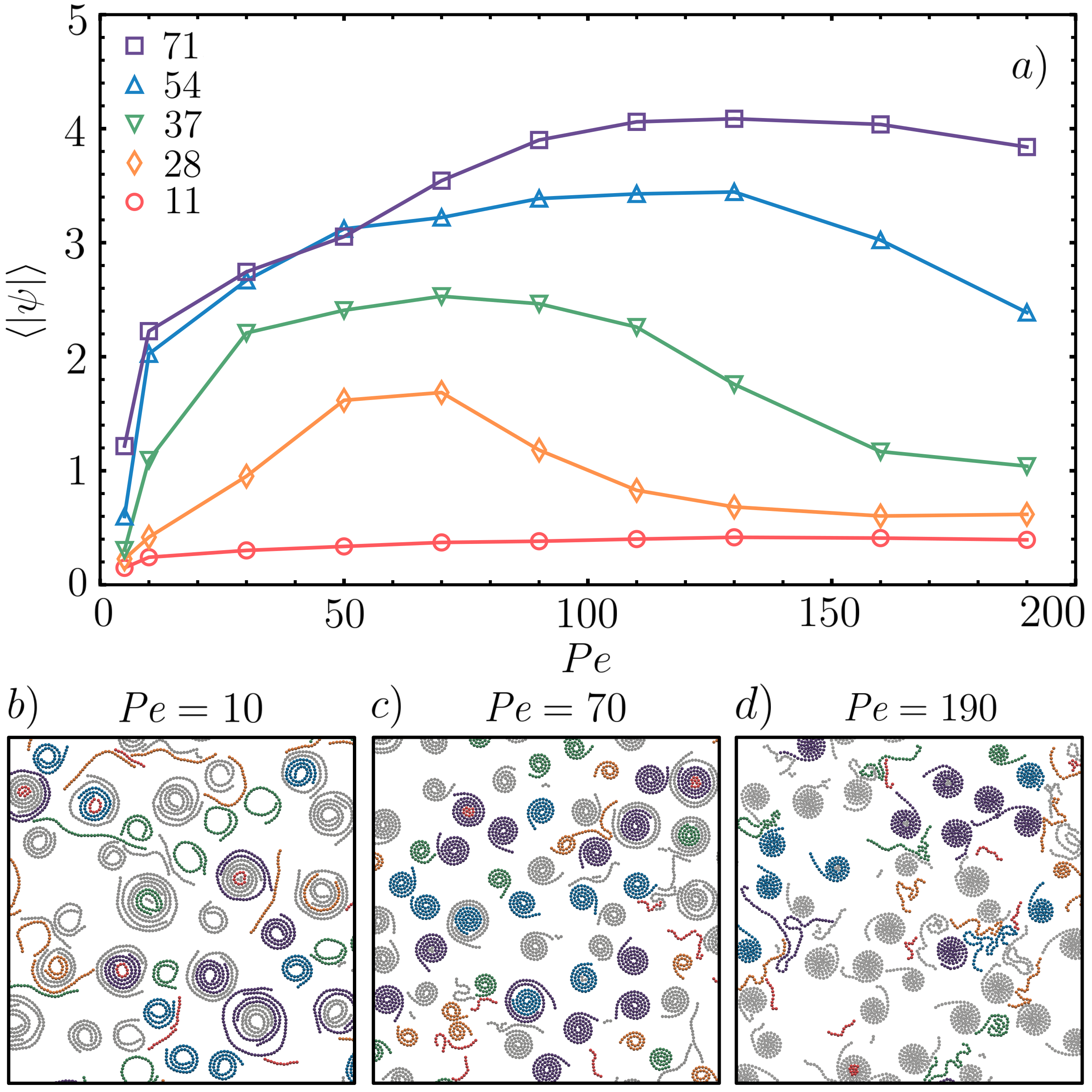}
    \caption{(a) Average turning number, $\left<|\psi|\right>$, as a function of the Péclet number for different filament lengths: red circles for $N_b = 11$, orange diamonds for $N_b = 28$, green down triangles for $N_b = 37$, blue up triangles for $N_b = 54$, and purple squares for $N_b = 71$. 
    For short filaments ($N_b = 11$), the turning number remains close to zero across all $Pe$ values, as the filaments are too short to form spiral structures.
    For longer filaments ($N_b \geq 28$), the turning number first increases and then decreases with $Pe$, indicating a reentrant behavior.
    The longer the filament, the higher the activity required to transition from an open-chain configuration to a tightly wound spiral configuration and back to an elongated one. (b-d) Snapshots of the system at steady-state: (b) at low ($Pe=10$), (c) intermediate ($Pe=70$), (d) and high Péclet number. Filaments are colored according to the legend in the top panel. Filaments in gray correspond to those whose length is not reported in the top panel.
    }
    \label{fig:tn}
\end{figure}

Fig. ~\ref{fig:tn}(a) shows the average turning number, $\left<|\psi|\right>$, as a function of the Péclet number for five representative filament lengths: one short ($N_b=11$), three medium ($N_b=28$, $N_b=37$ and $N_b=54$), and one long ($N_b=71$). Turning numbers for all other lengths are provided in the Supplementary Material~\cite{SI}.
Here, $\left<\cdot\right>$ denotes the ensemble average over all filaments of the same length.
Short filaments ($N_b = 11$) exhibit a turning number close to zero across all $Pe$, reflecting their inability to form spiral structures. 
In contrast, medium and long filaments ($N_b \geq 28$) display a non-monotonic trend: the turning number increases at intermediate $Pe$, indicating the formation of spiral structures, and decreases at high $Pe$, indicating a transition back to elongated configurations. 
This reentrant behavior is consistent with the monodisperse case~\cite{janzen2024density}.
Fig. ~\ref{fig:tn}(a) further indicates that longer filaments require higher activity to undergo this transition, showing that filaments of different lengths transition at different P\'eclet numbers.
Thus, polydispersity does not qualitatively alter the reentrant behavior, provided the filaments are sufficiently long to form spirals. However, for certain filament lengths, it shifts the P\'eclet number at which the transition occurs compared to the monodisperse case. Specifically, at the $Pe$ where filaments form stable spirals in the monodisperse system, the presence of shorter filaments already in the open-chain state perturbs the motion of longer filaments still in the spiral phase, promoting their premature unwinding (see Supplementary Material~\cite{SI}).
We also tested an exponentially decaying filament length distribution, inspired by self-assembled structures formed by active bifunctional Brownian particles~\cite{landi2024self}, and arguably more biologically relevant ~\cite{mitchison1984, howard2003, wegner1976, pollard2003}.
The results are qualitatively similar to those obtained with a uniform distribution (see Supplementary Material~\cite{SI}); therefore, we focus on the uniform case here.

The structural phases described above are illustrated in the system snapshots shown in Fig.~\ref{fig:tn}(b–d). At low $Pe$ (Fig.~\ref{fig:tn}(b)), filaments loosely wind into spiral structures. As $Pe$ increases (Fig.~\ref{fig:tn}(c)), these spirals become more tightly wound.
Finally, at high $Pe$ (Fig.~\ref{fig:tn}(d)), only long filaments form tightly wound spirals, while shorter filaments are fully unwound. Notably, Fig.~\ref{fig:tn}(b–d) also shows that, regardless of $Pe$, the system contains not only individual spiral filaments but also clusters of filaments that are interwound or wrapped around each other.

Such structures also appear in the monodisperse case~\cite{duman2018collective}; however, while in monodisperse systems they vanish at high $Pe$, they persist longer in the polydisperse case. This persistence arises from longer filaments, not yet uncoiled, winding around shorter ones (Fig.~\ref{fig:tn}(d)).
Eventually, at very high $Pe$, even these two-filament structures break down, as long filaments also uncoil. 
In the next section (Sec.~\ref{nested}), we characterize these `nested' structures using both structural and dynamical measurements.

\subsection{Nested Spirals as Emergent Collective Structures}
\label{nested}

The coiling behavior of filaments at intermediate activities (Fig.~\ref{fig:tn}) coexists with the formation of structurally organized assemblies, driven by filament–filament interactions. These assemblies consist of clusters of interlocking filaments arranged in spiral configurations, in agreement with the observations reported in Ref.\cite{duman2018collective}. 
We refer to these collective structures as nested spirals, defined as clusters in which the centers of mass of all constituent filaments lie within a threshold distance $\bar{r}$. The exact threshold value and a detailed explanation of the identification procedure are provided in the Supplementary Material~\cite{SI}.

Fig.~\ref{fig:formation_break} shows an example of how a nested spiral forms. The process begins with open-chain configurations (Fig.~\ref{fig:formation_break}(a)), and as activity drives motion, medium and long filaments begin to coil, initiating spiral formation (Fig.~\ref{fig:formation_break}(b)). During this stage, smaller filaments become trapped in the coiling structure, leading to the formation of a loose nested spiral (Fig.~\ref{fig:formation_break}(c)). This nested structure then transitions to a more stable configuration, in which filaments are more tightly wound (Fig.~\ref{fig:formation_break}(d)). However, depending on the activity, medium filaments may become unstable and uncoil (Fig.~\ref{fig:formation_break}(e)), ultimately leading to a two-filament nested spiral composed of one long and one short filament (Fig.~\ref{fig:formation_break}(f)). We anticipate that, at higher activities, even these two-filament structures eventually break down, as long filaments uncoil, and return to an open-chain state.

\begin{figure}[h!]
    \centering
    \includegraphics[width=\columnwidth]{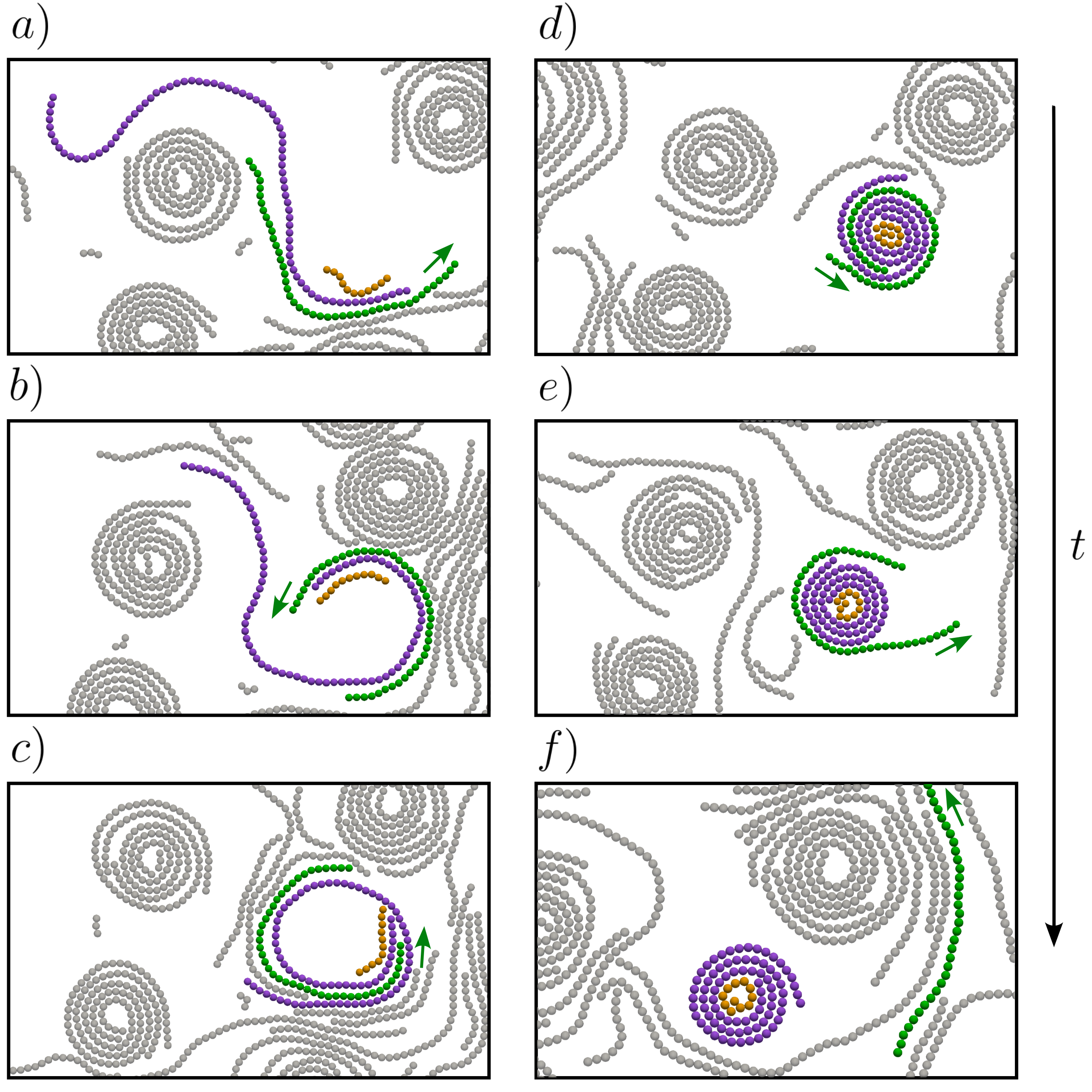}
    \caption{
    Snapshots illustrating the typical mechanism of formation (a–d) and transition to a two-filament state of a nested spiral (e–f). Snapshots show a short (orange), a medium (green), and a long (purple) filament during system equilibration at $Pe=10$.
    (a) Filaments start in open-chain configurations. (b) Driven by activity, the medium and long filaments begin to coil, initiating spiral formation. (c) As they coil, they capture and incorporate the smaller filament encountered along the way, forming a nested structure. (d) The nested structure becomes more compact, providing a relatively stable configuration for the long filament. (e) The medium filament becomes unstable and uncoils. (f) This results in a two-filament nested spiral composed of a long and a short filament.
    The green arrow in each panel indicates the self-propulsion direction of the medium filament.
    }
    \label{fig:formation_break}
\end{figure}

\begin{figure}[h!]
    \centering
    \includegraphics[width=\columnwidth]{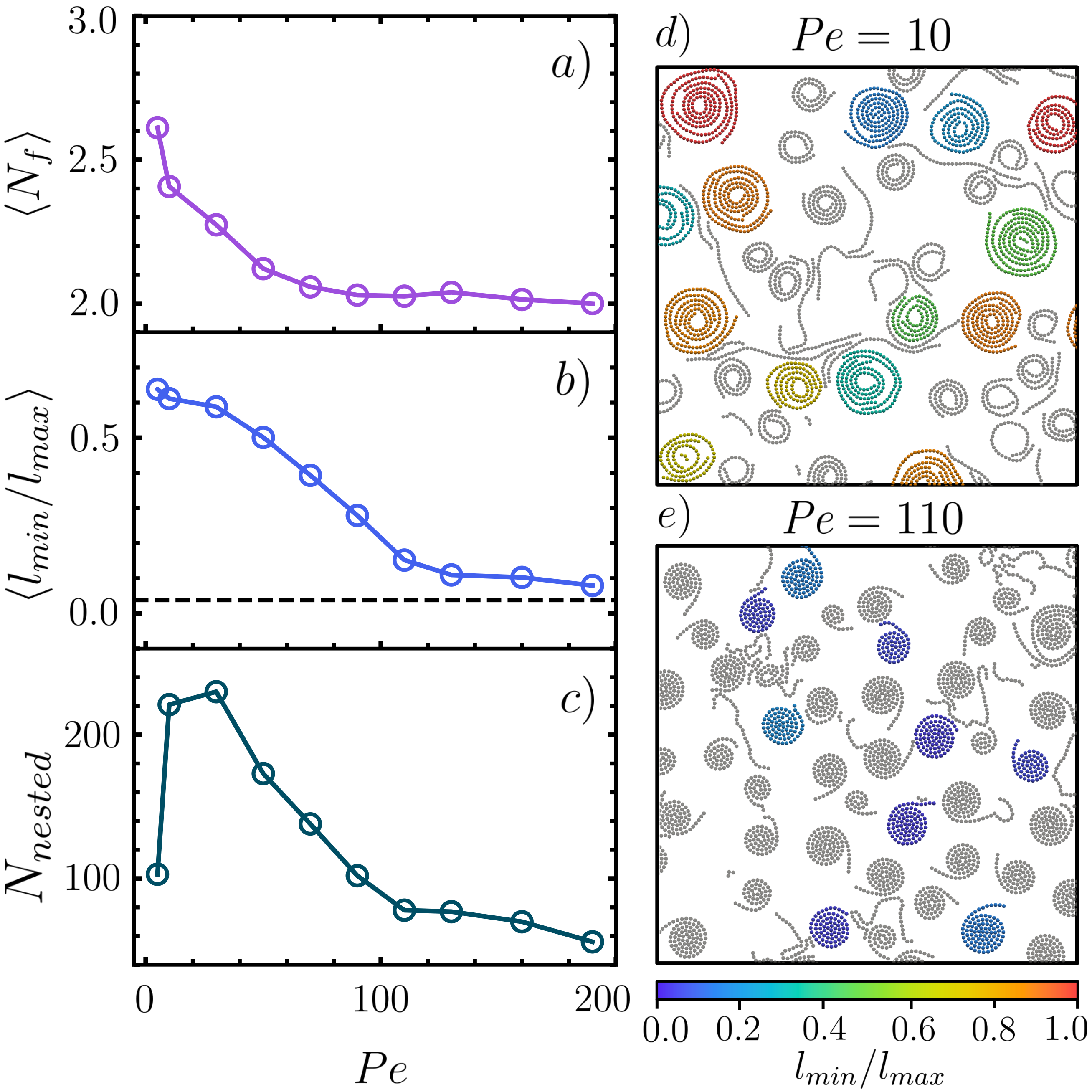}
    \caption{Structural properties of nested spirals as a function of the Péclet number. Panels (a, b) report ensemble-averaged quantities computed over all identified nested spirals. (a) Average number of filaments per nested spiral, $\langle N_f \rangle$. At low $Pe$, nested spirals typically contain more than two filaments, while at high $Pe$, they are predominantly composed of filament pairs.
    (b) Average length ratio of the shortest to the longest filament in each nested spiral, $\langle l_{\min}/l_{\max} \rangle$. The dotted black line at $y = 3/80$ denotes the minimum possible ratio based on the set of filament lengths in the system. At low $Pe$, filaments within nested spirals tend to be of similar length, while at high $Pe$, they exhibit pronounced length disparity. (c) Total number of nested spirals, $N_{\text{nested}}$. This quantity increases at low activity, reaches a peak, and then decreases with further increasing $Pe$. The decrease occurs in two stages: a sharp decay for $30 < Pe \leq 110$, followed by a more gradual reduction for $Pe > 110$. Initially, nested spirals form from the winding of medium and long filaments ($Pe \leq 30$). As $Pe$ increases, medium filaments begin to unwind, breaking apart multi-filament spirals. Concurrently, long filaments wrap around short filaments, forming predominantly two-filament nested spirals. At high $Pe$, these structures dissolve as long filaments themselves unwind. (d, e) Representative snapshots of the system in steady state at $Pe = 10$ (d) and $Pe = 110$ (e). Filaments that belong to nested spirals are colored according to the length ratio $l_{\min}/l_{\max}$ of the corresponding structure. Filaments not participating in nested spirals are shown in gray.
    }
    \label{fig:transition_nested}
\end{figure}

The structure and composition of nested spirals vary markedly with increasing activity. 
We begin by examining the average number of filaments per nested spiral, $\langle N_f \rangle$. As shown in Fig.~\ref{fig:transition_nested}(a), at low $Pe$, nested spirals typically comprise more than two filaments, while at high $Pe$, the average converges to two, indicating that they ultimately consist of filament pairs.
This trend is further supported by the probability distributions of the number of filaments per nested spiral across different $Pe$ (see Supplementary Material \cite{SI}).
Moreover, the filament lengths involved in these nested structures change with the P\'eclet number. 
This effect is quantified by measuring the average ratio of the shortest, $l_{min}$, to the longest filament, $l_{max}$, in each structure, $\langle l_{\min}/l_{\max} \rangle$ (Fig.~\ref{fig:transition_nested}(b)).
At low $Pe$, this ratio is close to one, indicating that nested spirals are typically composed of filaments of similar length. As $Pe$ increases, this ratio decreases toward its minimum, reflecting a transition to structures formed by one short and one long filament. What we observe here is consistent with the monodisperse case~\cite{duman2018collective}: as $Pe$ increases, nested spirals composed of filaments of similar lengths progressively disappear.
Additionally, long filaments dominate at low $Pe$, as indicated by the increased average filament length in this regime (see Supplementary Material~\cite{SI}).
Overall, as the Péclet number increases, nested spirals transition from a state in which they typically consist of multiple long filaments to one where they are composed of only two filaments, one long and one short.

Finally, we observe that the number of nested spirals, $N_{nested}$, initially increases with activity and then decreases, as shown in Fig.~\ref{fig:transition_nested}(c). 
This decrease exhibits a two-stage behavior: a rapid initial decay followed by a more gradual reduction, reflecting the formation and disruption mechanisms of nested spirals described earlier (Fig.~\ref{fig:formation_break}). 
At low $Pe$, multi-filament nested spirals are common, as medium and long filaments tend to form coiled structures (Fig.~\ref{fig:transition_nested}(d)). As $Pe$ increases, medium filaments progressively unwind, destabilizing these structures. This results in a transient regime where two-filament nested spirals dominate, typically formed by a long filament wrapping around a short one (Fig.~\ref{fig:transition_nested}(e)). At even higher $Pe$, long filaments also unwind, leading to the eventual breakdown of nested spirals. 

\begin{figure}[h!]
    \centering
    \includegraphics[width=\columnwidth]{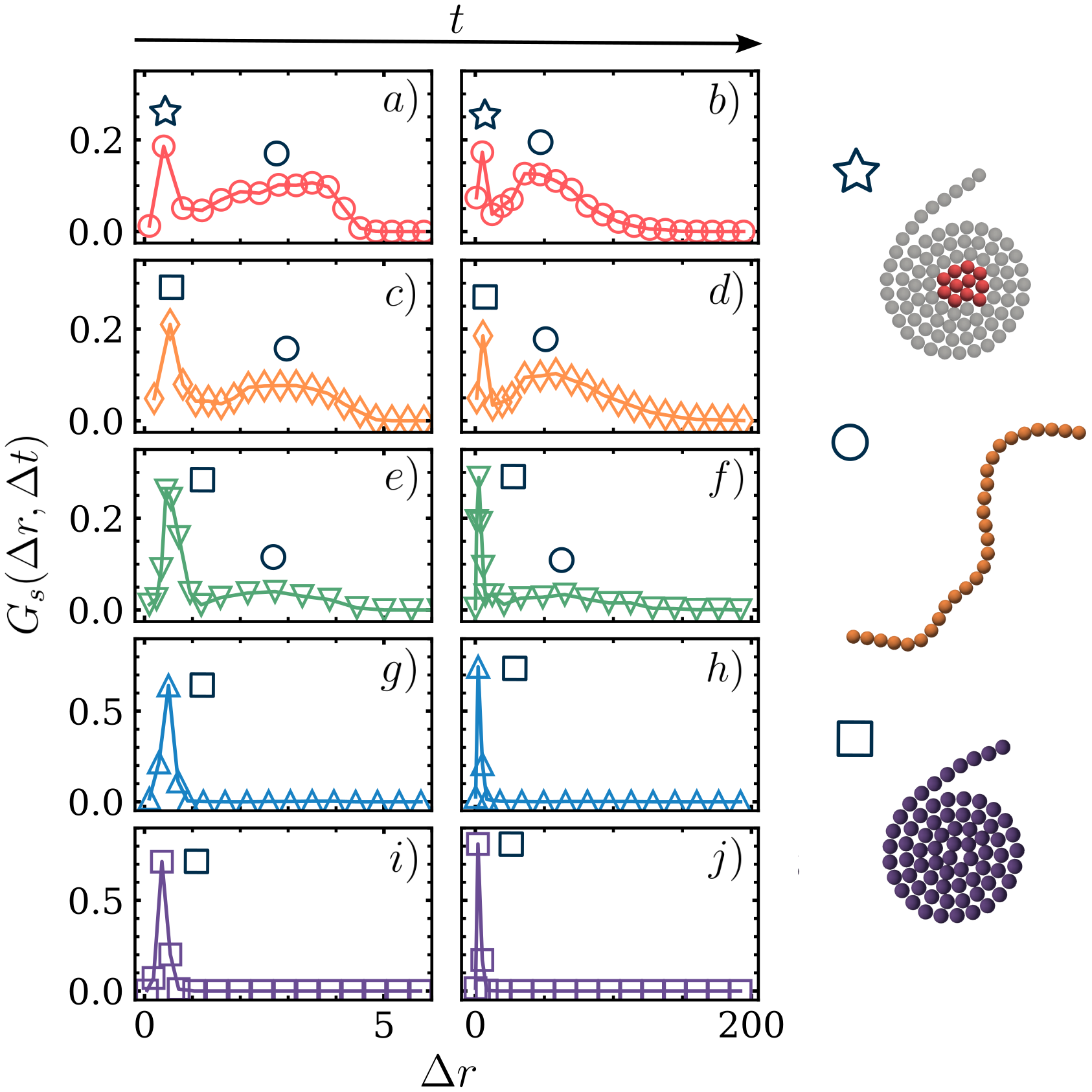}
    \caption{van Hove distribution for filaments of different lengths at $Pe = 110$. Each pair of panels corresponds to a filament length: red circles for $N_b = 11$ (a, b), orange diamonds for $N_b = 28$ (c, d), green downward triangles for $N_b = 37$ (e, f), blue upward triangles for $N_b = 54$ (g, h), and purple squares for $N_b = 71$ (i, j). Left panels (a, c, e, g, i) correspond to a short time, $\Delta t = 1.5$, while right panels (b, d, f, h, j) correspond to a longer time, $\Delta t = 150$. (a, b) The distribution for short filaments ($l = 11$) displays a clear bimodal structure, indicating the presence of two distinct dynamical populations: one corresponding to filaments in elongated, motile configurations and another to filaments confined within nested spirals. When encapsulated by longer filaments, short filaments inherit the constrained dynamics of the nested structure. (c - f) Medium-length filaments ($N_b = 28$, $N_b = 37$) also exhibit bimodal distributions, reflecting their coexistence in both spiral and elongated states. At this activity, medium filaments begin to unwind, resulting in a mixture of dynamical behaviors within the population. (g - j) For long filaments ($N_b = 54$, $N_b = 71$), the distribution shows a single peak, consistent with a population dominated by spiral conformations. At $Pe = 110$, these filaments remain wound and exhibit a single, confined dynamical behavior, whether in isolated spirals or nested structures.
    }
    \label{fig:displacements}
\end{figure}

The structural reorganization of filaments into nested spirals has a strong impact on their dynamics. In particular, the formation of nested spirals gives rise to distinct displacement behaviors that depend on filament length and structural organization. To quantify these effects, we analyze the self-part of the van Hove distribution~\cite{hansen1986theory, helfferich2018}, which describes the probability distribution of filament displacements. Specifically, it measures how far the center of mass of each filament has moved after a time lag $\Delta t$.
The van Hove distribution is defined as
\begin{equation}
G_s(\Delta r, \Delta t) = \left\langle \frac{1}{N_f^{\,\text{tot}}} \sum_{i=1}^{N_f^{\,\text{tot}}} \delta\left( \Delta r - \left| \mathbf{r}^{\,\text{cm}}_i(t+\Delta t) - \mathbf{r}^{\,\text{cm}}_i(t) \right| \right) \right\rangle,
\end{equation}
where $N_f^{\,\text{tot}}$ is the total number of filaments in the system, $\mathbf{r}^{\,\text{cm}}_i(t)$ denotes the position of the center of mass of filament $i$ at time $t$, $\Delta r$ represents the magnitude of the displacement between time $t$ and time $t+\Delta t$, and $\delta(\ldots)$ is the Dirac delta function. 
The time average $\langle \cdot \rangle$ is computed over a single simulation trajectory using $t = n \Delta t$, with $n = 0, 1, 2, 3, 4,$ and $5$, after equilibration. In addition, we average over 10 independent simulations with different random seeds.
We compute this distribution for different filament lengths at $Pe = 110$ and show these values in Fig. ~\ref{fig:displacements} for two representative lag times: a short lag time, $\Delta t = 1.5$, and a longer one, $\Delta t = 150$. The Péclet number is set to $Pe = 110$, corresponding to an activity at which the transition to two-filament nested spirals has just occurred (Fig.~\ref{fig:transition_nested}).

Short filaments (Fig.~\ref{fig:displacements}(a,b)) exhibit a bimodal structure at both $\Delta t = 1.5$ and $\Delta t = 150$, indicating the coexistence of two distinct dynamic behaviors. This is reflected in the presence of two peaks in the displacement distribution: one at smaller $\Delta r$ and one at larger $\Delta r$. The peak at larger $\Delta r$ corresponds to filaments in elongated configurations undergoing directed motion, while the peak at smaller $\Delta r$ arises from filaments incorporated into nested spiral structures. These confined filaments behave like spirals and exhibit limited mobility, resulting in smaller displacements.
A similar bimodal distribution is observed for medium filaments (Fig.~\ref{fig:displacements}(c-f)), but the origin of this behavior is different. In this case, the filament length is sufficient to form spirals, but at this P\'eclet number some spirals, whether single or nested, begin to unwind. As a result, medium filaments can be found in either spiral or open-chain configurations, each associated with distinct displacement statistics. 
In contrast, the distribution for long filaments (Fig.~\ref{fig:displacements}(g-j)) displays a single peak, consistent with the fact that these filaments are mainly in tightly wound spiral conformations either as one filament spirals or part of nested spirals. In both cases, their dynamics is similarly constrained and do not exhibit the fast, directed motion characteristic of open-chains.
Across all filament lengths, we observe consistent dynamical behavior at both short (Fig.~\ref{fig:displacements}(a, c, e, g, i)) and long times (Fig.~\ref{fig:displacements}(b, d, f, h, j)), indicating that the underlying dynamics is robust over time.

\section{Conclusions}
In summary, our work shows that while a system composed of a distribution of polymer lengths behaves qualitatively similarly to a monodisperse system, some differences still emerge. 
Understanding how structural diversity influences collective behavior in active matter is key to bridging idealized models and real-world systems, where heterogeneity is often unavoidable. In this work, we investigated how filament length polydispersity affects the self-organization of active semiflexible filaments. Our work demonstrates that while nested spiral structures can form in both monodisperse and polydisperse systems, polydispersity profoundly alters their stability and organization, promoting hierarchical wrapping and persistent confinement at high activity. These effects arise from dynamic interactions across length scales: long filaments can trap and stabilize shorter ones, extending the lifetime and robustness of spiral assemblies beyond what is observed in monodisperse populations.

In particular, by varying the P\'eclet number, we observe the same structural phases seen in the monodisperse case~\cite{IseleHolder2015,Prathyusha2018,duman2018collective,janzen2024density}: a polymer melt phase, a transition to a pure spiral phase at intermediate $Pe$, and reentrant unwinding of spirals at high $Pe$. Thus, polydispersity does not fundamentally alter the reentrant behavior, provided filaments are sufficiently long to form spirals. However, it can shift the P\'eclet number at which these transitions occur. Specifically, at the $Pe$ where spirals are stable in the monodisperse case, shorter filaments already in the open-chain state can perturb the motion of longer spirals, promoting their premature unwinding.

Moreover, in monodisperse systems, it has been shown that at intermediate $Pe$, filaments can form interlocked or wrapped configurations, i.e., nested spirals. However, these structures disappear as $Pe$ increases \cite{duman2018collective}. In the polydisperse case, by contrast, such nested structures persist even at high Péclet numbers. As in the monodisperse case, nested spirals composed of filaments of comparable length vanish at high $Pe$. In the very high $Pe$ regime, the most common nested spirals consist of a long filament wrapped around a much shorter one. This occurs because short filaments become trapped inside the tightly coiled spirals. We anticipate that at even higher $Pe$, these nested structures will eventually break down as the long filaments themselves uncoil and transition to open-chain configurations.

Finally, we analyzed the dynamics of nested spirals with the van Hove distribution. For short filaments the distribution is bimodal, revealing two distinct dynamical behaviors: filaments trapped inside nested spirals show small displacements characteristic of confined spirals, whereas open-chain filaments move more freely and display larger displacement. Medium filaments exhibit a similar bimodal profile because, at this P\'eclet number, they begin to unwind and can exist either in spiral (single or nested) or open-chain configurations. In contrast, long filaments remain almost exclusively in tightly wound spiral conformations, leading to a single narrow peak in the van Hove distribution that reflects their constrained dynamics.

A potential extension of this work would be to introduce length-dependent propulsion, allowing different filament lengths to self-propel at different speeds, better reflecting conditions in biological systems \cite{landi2024self} and further clarifying the role of polydispersity in active filament systems.

\section*{Acknowledgements}
DMF and GJ acknowledge support from the Comunidad de Madrid and the Complutense University of Madrid (Spain) through the Atracción de Talento program (Grant No. 2022-T1/TIC-24007). DMF also acknowledges support from MINECO (Grant No. PID2023-148991NA-I00. C.V. acknowledges fund-
ings IHRC22/00002 and PID2022-140407NB-C21 from MINECO. 

\balance

\bibliography{biblo}

\providecommand{\noopsort}[1]{}\providecommand{\singleletter}[1]{#1}%
\providecommand*{\mcitethebibliography}{\thebibliography}
\csname @ifundefined\endcsname{endmcitethebibliography}
{\let\endmcitethebibliography\endthebibliography}{}
\begin{mcitethebibliography}{52}
\providecommand*{\natexlab}[1]{#1}
\providecommand*{\mciteSetBstSublistMode}[1]{}
\providecommand*{\mciteSetBstMaxWidthForm}[2]{}
\providecommand*{\mciteBstWouldAddEndPuncttrue}
  {\def\EndOfBibitem{\unskip.}}
\providecommand*{\mciteBstWouldAddEndPunctfalse}
  {\let\EndOfBibitem\relax}
\providecommand*{\mciteSetBstMidEndSepPunct}[3]{}
\providecommand*{\mciteSetBstSublistLabelBeginEnd}[3]{}
\providecommand*{\EndOfBibitem}{}
\mciteSetBstSublistMode{f}
\mciteSetBstMaxWidthForm{subitem}
{(\emph{\alph{mcitesubitemcount}})}
\mciteSetBstSublistLabelBeginEnd{\mcitemaxwidthsubitemform\space}
{\relax}{\relax}

\bibitem[te~Vrugt and Wittkowski(2025)]{metareview}
M.~te~Vrugt and R.~Wittkowski, \emph{The European Physical Journal E}, 2025,
  \textbf{48}, 12\relax
\mciteBstWouldAddEndPuncttrue
\mciteSetBstMidEndSepPunct{\mcitedefaultmidpunct}
{\mcitedefaultendpunct}{\mcitedefaultseppunct}\relax
\EndOfBibitem
\bibitem[Sciortino \emph{et~al.}(2025)Sciortino, Faizi,
  Fedosov,\emph{et~al.}]{alfredo_new}
A.~Sciortino, H.~A. Faizi, D.~A. Fedosov \emph{et~al.}, \emph{Nature Physics},
  2025, \textbf{21}, 799--807\relax
\mciteBstWouldAddEndPuncttrue
\mciteSetBstMidEndSepPunct{\mcitedefaultmidpunct}
{\mcitedefaultendpunct}{\mcitedefaultseppunct}\relax
\EndOfBibitem
\bibitem[Ganguly \emph{et~al.}(2012)Ganguly, Williams, Palacios, and
  Goldstein]{ganguly2012cytoplasmic}
S.~Ganguly, L.~S. Williams, I.~M. Palacios and R.~E. Goldstein,
  \emph{Proceedings of the National Academy of Sciences}, 2012, \textbf{109},
  15109--15114\relax
\mciteBstWouldAddEndPuncttrue
\mciteSetBstMidEndSepPunct{\mcitedefaultmidpunct}
{\mcitedefaultendpunct}{\mcitedefaultseppunct}\relax
\EndOfBibitem
\bibitem[Yaman \emph{et~al.}(2019)Yaman, Demir, Vetter, and Kocabas]{Yaman2019}
Y.~I. Yaman, E.~Demir, R.~Vetter and A.~Kocabas, \emph{Nature communications},
  2019, \textbf{10}, 1--9\relax
\mciteBstWouldAddEndPuncttrue
\mciteSetBstMidEndSepPunct{\mcitedefaultmidpunct}
{\mcitedefaultendpunct}{\mcitedefaultseppunct}\relax
\EndOfBibitem
\bibitem[Ding \emph{et~al.}(2019)Ding, Schumacher, Javer, Endres, and
  Brown]{serena2019}
S.~S. Ding, L.~J. Schumacher, A.~E. Javer, R.~G. Endres and A.~E. Brown,
  \emph{eLife}, 2019, \textbf{8}, e43318\relax
\mciteBstWouldAddEndPuncttrue
\mciteSetBstMidEndSepPunct{\mcitedefaultmidpunct}
{\mcitedefaultendpunct}{\mcitedefaultseppunct}\relax
\EndOfBibitem
\bibitem[Auer \emph{et~al.}(2019)Auer, Oliver, Rajendram, Lin, Yao, Jensen, and
  Weibel]{auer_bacterial_2019}
G.~K. Auer, P.~M. Oliver, M.~Rajendram, T.-Y. Lin, Q.~Yao, G.~J. Jensen and
  D.~B. Weibel, \emph{mBio}, 2019, \textbf{10}, e00210‑--19\relax
\mciteBstWouldAddEndPuncttrue
\mciteSetBstMidEndSepPunct{\mcitedefaultmidpunct}
{\mcitedefaultendpunct}{\mcitedefaultseppunct}\relax
\EndOfBibitem
\bibitem[Deblais \emph{et~al.}(2020)Deblais, Maggs, Bonn, and
  Woutersen]{deblais2020phase}
A.~Deblais, A.~Maggs, D.~Bonn and S.~Woutersen, \emph{Physical Review Letters},
  2020, \textbf{124}, 208006\relax
\mciteBstWouldAddEndPuncttrue
\mciteSetBstMidEndSepPunct{\mcitedefaultmidpunct}
{\mcitedefaultendpunct}{\mcitedefaultseppunct}\relax
\EndOfBibitem
\bibitem[Deblais \emph{et~al.}(2020)Deblais, Woutersen, and
  Bonn]{deblais2020rheology}
A.~Deblais, S.~Woutersen and D.~Bonn, \emph{Physical Review Letters}, 2020,
  \textbf{124}, 188002\relax
\mciteBstWouldAddEndPuncttrue
\mciteSetBstMidEndSepPunct{\mcitedefaultmidpunct}
{\mcitedefaultendpunct}{\mcitedefaultseppunct}\relax
\EndOfBibitem
\bibitem[Nguyen \emph{et~al.}(2021)Nguyen, Ozkan-Aydin, Tuazon, Goldman,
  Bhamla, and Peleg]{nguyen2021emergent}
C.~Nguyen, Y.~Ozkan-Aydin, H.~Tuazon, D.~I. Goldman, M.~S. Bhamla and O.~Peleg,
  \emph{Frontiers in Physics}, 2021, \textbf{9}, 734499\relax
\mciteBstWouldAddEndPuncttrue
\mciteSetBstMidEndSepPunct{\mcitedefaultmidpunct}
{\mcitedefaultendpunct}{\mcitedefaultseppunct}\relax
\EndOfBibitem
\bibitem[Yan \emph{et~al.}(2016)Yan, Han, Zhang, Xu, Luijten, and
  Granick]{yan_2016}
J.~Yan, M.~Han, J.~Zhang, C.~Xu, E.~Luijten and S.~Granick, \emph{Nature
  Materials}, 2016, \textbf{15}, 1095--1099\relax
\mciteBstWouldAddEndPuncttrue
\mciteSetBstMidEndSepPunct{\mcitedefaultmidpunct}
{\mcitedefaultendpunct}{\mcitedefaultseppunct}\relax
\EndOfBibitem
\bibitem[Wen \emph{et~al.}(2012)Wen, Zheng, Li, Li, Sun, and
  Shi]{wen2012polymerlike}
P.-P. Wen, N.~Zheng, L.-S. Li, H.~Li, G.~Sun and Q.-F. Shi, \emph{Physical
  Review E}, 2012, \textbf{85}, 031301\relax
\mciteBstWouldAddEndPuncttrue
\mciteSetBstMidEndSepPunct{\mcitedefaultmidpunct}
{\mcitedefaultendpunct}{\mcitedefaultseppunct}\relax
\EndOfBibitem
\bibitem[Soh \emph{et~al.}(2019)Soh, Gengaro, Klotz, and Doyle]{soh2019self}
B.~W. Soh, I.~R. Gengaro, A.~R. Klotz and P.~S. Doyle, \emph{Physical Review
  Research}, 2019, \textbf{1}, 033194\relax
\mciteBstWouldAddEndPuncttrue
\mciteSetBstMidEndSepPunct{\mcitedefaultmidpunct}
{\mcitedefaultendpunct}{\mcitedefaultseppunct}\relax
\EndOfBibitem
\bibitem[Marvi \emph{et~al.}(2014)Marvi, Gong, Gravish, Astley, Travers,
  Hatton, Mendelson~III, Choset, Hu, and Goldman]{marvi2014sidewinding}
H.~Marvi, C.~Gong, N.~Gravish, H.~Astley, M.~Travers, R.~L. Hatton, J.~R.
  Mendelson~III, H.~Choset, D.~L. Hu and D.~I. Goldman, \emph{Science}, 2014,
  \textbf{346}, 224--229\relax
\mciteBstWouldAddEndPuncttrue
\mciteSetBstMidEndSepPunct{\mcitedefaultmidpunct}
{\mcitedefaultendpunct}{\mcitedefaultseppunct}\relax
\EndOfBibitem
\bibitem[Ozkan-Aydin \emph{et~al.}(2021)Ozkan-Aydin, Goldman, and
  Bhamla]{ozkan2021collective}
Y.~Ozkan-Aydin, D.~I. Goldman and M.~S. Bhamla, \emph{Proceedings of the
  National Academy of Sciences}, 2021, \textbf{118}, e2010542118\relax
\mciteBstWouldAddEndPuncttrue
\mciteSetBstMidEndSepPunct{\mcitedefaultmidpunct}
{\mcitedefaultendpunct}{\mcitedefaultseppunct}\relax
\EndOfBibitem
\bibitem[Bourdieu \emph{et~al.}(1995)Bourdieu, Duke, Elowitz, Winkelmann,
  Leibler, and Libchaber]{bourdieu_spiral_1995}
L.~Bourdieu, T.~Duke, M.~B. Elowitz, D.~A. Winkelmann, S.~Leibler and
  A.~Libchaber, \emph{Phys. Rev. Lett.}, 1995, \textbf{75}, 176--179\relax
\mciteBstWouldAddEndPuncttrue
\mciteSetBstMidEndSepPunct{\mcitedefaultmidpunct}
{\mcitedefaultendpunct}{\mcitedefaultseppunct}\relax
\EndOfBibitem
\bibitem[Schaller \emph{et~al.}(2010)Schaller, Weber, Semmrich, Frey, and
  Bausch]{Schaller2010}
V.~Schaller, C.~Weber, C.~Semmrich, E.~Frey and A.~R. Bausch, \emph{Nature},
  2010, \textbf{467}, 73--77\relax
\mciteBstWouldAddEndPuncttrue
\mciteSetBstMidEndSepPunct{\mcitedefaultmidpunct}
{\mcitedefaultendpunct}{\mcitedefaultseppunct}\relax
\EndOfBibitem
\bibitem[Sumino \emph{et~al.}(2012)Sumino, Nagai, Shitaka, Tanaka, Yoshikawa,
  Chaté, and Oiwa]{sumino_2012}
Y.~Sumino, K.~H. Nagai, Y.~Shitaka, D.~Tanaka, K.~Yoshikawa, H.~Chaté and
  K.~Oiwa, \emph{Nature}, 2012, \textbf{483}, 448--452\relax
\mciteBstWouldAddEndPuncttrue
\mciteSetBstMidEndSepPunct{\mcitedefaultmidpunct}
{\mcitedefaultendpunct}{\mcitedefaultseppunct}\relax
\EndOfBibitem
\bibitem[Sciortino and Bausch(2021)]{sciortino_2021}
A.~Sciortino and A.~R. Bausch, \emph{Proceedings of the National Academy of
  Sciences}, 2021, \textbf{118}, e2017047118\relax
\mciteBstWouldAddEndPuncttrue
\mciteSetBstMidEndSepPunct{\mcitedefaultmidpunct}
{\mcitedefaultendpunct}{\mcitedefaultseppunct}\relax
\EndOfBibitem
\bibitem[Chan \emph{et~al.}(2007)Chan, Calder, Fox, and
  Lloyd]{cortical_microtubule}
J.~Chan, G.~Calder, S.~Fox and C.~Lloyd, \emph{Nature cell biology}, 2007,
  \textbf{9}, 171--5\relax
\mciteBstWouldAddEndPuncttrue
\mciteSetBstMidEndSepPunct{\mcitedefaultmidpunct}
{\mcitedefaultendpunct}{\mcitedefaultseppunct}\relax
\EndOfBibitem
\bibitem[Lin \emph{et~al.}(2014)Lin, Lo, and Lo]{lin_2014}
S.-N. Lin, W.-C. Lo and C.-J. Lo, \emph{Soft Matter}, 2014, \textbf{10},
  760--766\relax
\mciteBstWouldAddEndPuncttrue
\mciteSetBstMidEndSepPunct{\mcitedefaultmidpunct}
{\mcitedefaultendpunct}{\mcitedefaultseppunct}\relax
\EndOfBibitem
\bibitem[Isele-Holder \emph{et~al.}(2015)Isele-Holder, Elgeti, and
  Gompper]{IseleHolder2015}
R.~E. Isele-Holder, J.~Elgeti and G.~Gompper, \emph{Soft Matter}, 2015,
  \textbf{11}, 7181--7190\relax
\mciteBstWouldAddEndPuncttrue
\mciteSetBstMidEndSepPunct{\mcitedefaultmidpunct}
{\mcitedefaultendpunct}{\mcitedefaultseppunct}\relax
\EndOfBibitem
\bibitem[Duman \emph{et~al.}(2018)Duman, Isele-Holder, Elgeti, and
  Gompper]{duman2018collective}
O.~Duman, R.~E. Isele-Holder, J.~Elgeti and G.~Gompper, \emph{Soft matter},
  2018, \textbf{14}, 4483--4494\relax
\mciteBstWouldAddEndPuncttrue
\mciteSetBstMidEndSepPunct{\mcitedefaultmidpunct}
{\mcitedefaultendpunct}{\mcitedefaultseppunct}\relax
\EndOfBibitem
\bibitem[Janzen and Matoz-Fernandez(2024)]{janzen2024density}
G.~Janzen and D.~A. Matoz-Fernandez, \emph{Soft Matter}, 2024, \textbf{20},
  6618--6626\relax
\mciteBstWouldAddEndPuncttrue
\mciteSetBstMidEndSepPunct{\mcitedefaultmidpunct}
{\mcitedefaultendpunct}{\mcitedefaultseppunct}\relax
\EndOfBibitem
\bibitem[Prathyusha \emph{et~al.}(2018)Prathyusha, Henkes, and
  Sknepnek]{Prathyusha2018}
K.~R. Prathyusha, S.~Henkes and R.~Sknepnek, \emph{Physical Review E}, 2018,
  \textbf{97}, 022606\relax
\mciteBstWouldAddEndPuncttrue
\mciteSetBstMidEndSepPunct{\mcitedefaultmidpunct}
{\mcitedefaultendpunct}{\mcitedefaultseppunct}\relax
\EndOfBibitem
\bibitem[Shee \emph{et~al.}(2021)Shee, Gupta, Chaudhuri, and
  Chaudhuri]{Shee2021}
A.~Shee, N.~Gupta, A.~Chaudhuri and D.~Chaudhuri, \emph{Soft Matter}, 2021,
  \textbf{17}, 2120--2131\relax
\mciteBstWouldAddEndPuncttrue
\mciteSetBstMidEndSepPunct{\mcitedefaultmidpunct}
{\mcitedefaultendpunct}{\mcitedefaultseppunct}\relax
\EndOfBibitem
\bibitem[Anand(2025)]{anand_2025}
S.~K. Anand, \emph{Journal of Physics: Condensed Matter}, 2025, \textbf{37},
  185101\relax
\mciteBstWouldAddEndPuncttrue
\mciteSetBstMidEndSepPunct{\mcitedefaultmidpunct}
{\mcitedefaultendpunct}{\mcitedefaultseppunct}\relax
\EndOfBibitem
\bibitem[Bianco \emph{et~al.}(2018)Bianco, Locatelli, and
  Malgaretti]{Bianco2108}
V.~Bianco, E.~Locatelli and P.~Malgaretti, \emph{Phys. Rev. Lett.}, 2018,
  \textbf{121}, 217802\relax
\mciteBstWouldAddEndPuncttrue
\mciteSetBstMidEndSepPunct{\mcitedefaultmidpunct}
{\mcitedefaultendpunct}{\mcitedefaultseppunct}\relax
\EndOfBibitem
\bibitem[Li \emph{et~al.}(2023)Li, Wu, Hao, Lei, and Ma]{Li2023}
J.-X. Li, S.~Wu, L.-L. Hao, Q.-L. Lei and Y.-Q. Ma, \emph{Phys. Rev. Res.},
  2023, \textbf{5}, 043064\relax
\mciteBstWouldAddEndPuncttrue
\mciteSetBstMidEndSepPunct{\mcitedefaultmidpunct}
{\mcitedefaultendpunct}{\mcitedefaultseppunct}\relax
\EndOfBibitem
\bibitem[Winkler and Gompper(2020)]{winkler_physics_2020}
R.~G. Winkler and G.~Gompper, \emph{The Journal of Chemical Physics}, 2020,
  \textbf{153}, 040901\relax
\mciteBstWouldAddEndPuncttrue
\mciteSetBstMidEndSepPunct{\mcitedefaultmidpunct}
{\mcitedefaultendpunct}{\mcitedefaultseppunct}\relax
\EndOfBibitem
\bibitem[Gopinathan \emph{et~al.}(2007)Gopinathan, Lee, Schwarz, and
  Liu]{gopinathan_2007}
A.~Gopinathan, K.-C. Lee, J.~M. Schwarz and A.~J. Liu, \emph{Physical Review
  Letters}, 2007, \textbf{99}, 058103\relax
\mciteBstWouldAddEndPuncttrue
\mciteSetBstMidEndSepPunct{\mcitedefaultmidpunct}
{\mcitedefaultendpunct}{\mcitedefaultseppunct}\relax
\EndOfBibitem
\bibitem[Pavlov \emph{et~al.}(2007)Pavlov, Muhlrad, Cooper, Wear, and
  Reisler]{pavlov_2007}
D.~Pavlov, A.~Muhlrad, J.~Cooper, M.~Wear and E.~Reisler, \emph{Journal of
  molecular biology}, 2007, \textbf{365}, 1350--1358\relax
\mciteBstWouldAddEndPuncttrue
\mciteSetBstMidEndSepPunct{\mcitedefaultmidpunct}
{\mcitedefaultendpunct}{\mcitedefaultseppunct}\relax
\EndOfBibitem
\bibitem[Humphrey \emph{et~al.}(2002)Humphrey, Duggan, Saha, Smith, and
  Käs]{humphrey_2002}
D.~Humphrey, C.~Duggan, D.~Saha, D.~Smith and J.~Käs, \emph{Nature}, 2002,
  \textbf{416}, 413--416\relax
\mciteBstWouldAddEndPuncttrue
\mciteSetBstMidEndSepPunct{\mcitedefaultmidpunct}
{\mcitedefaultendpunct}{\mcitedefaultseppunct}\relax
\EndOfBibitem
\bibitem[Lietor-Santos \emph{et~al.}(2010)Lietor-Santos, Kim, Lynch,
  Fernandez-Nieves, and Weitz]{lietor_2010}
J.~J. Lietor-Santos, C.~Kim, M.~L. Lynch, A.~Fernandez-Nieves and D.~A. Weitz,
  \emph{Langmuir}, 2010, \textbf{26}, 3174--3178\relax
\mciteBstWouldAddEndPuncttrue
\mciteSetBstMidEndSepPunct{\mcitedefaultmidpunct}
{\mcitedefaultendpunct}{\mcitedefaultseppunct}\relax
\EndOfBibitem
\bibitem[Kumar \emph{et~al.}(2024)Kumar, Murali, Subramaniam, Singh, and
  Thutupalli]{kumar_2024}
M.~Kumar, A.~Murali, A.~G. Subramaniam, R.~Singh and S.~Thutupalli,
  \emph{Nature Communications}, 2024, \textbf{15}, 4903\relax
\mciteBstWouldAddEndPuncttrue
\mciteSetBstMidEndSepPunct{\mcitedefaultmidpunct}
{\mcitedefaultendpunct}{\mcitedefaultseppunct}\relax
\EndOfBibitem
\bibitem[Li \emph{et~al.}(2025)Li, Zhang, Zhang, Wang, Zhang, Ding, and
  Han]{Li2025Polydispersity}
J.~Li, C.~Zhang, Q.~Zhang, S.~Wang, R.~Zhang, Z.~Ding and Y.~Han,
  \emph{Macromolecules}, 2025, \textbf{58}, 3208--3220\relax
\mciteBstWouldAddEndPuncttrue
\mciteSetBstMidEndSepPunct{\mcitedefaultmidpunct}
{\mcitedefaultendpunct}{\mcitedefaultseppunct}\relax
\EndOfBibitem
\bibitem[De~Filippo \emph{et~al.}(2023)De~Filippo, Del~Galdo, Corsi,
  De~Michele, and Capone]{de_filippo2023}
C.~A. De~Filippo, S.~Del~Galdo, P.~Corsi, C.~De~Michele and B.~Capone,
  \emph{Soft Matter}, 2023, \textbf{19}, 1732--1738\relax
\mciteBstWouldAddEndPuncttrue
\mciteSetBstMidEndSepPunct{\mcitedefaultmidpunct}
{\mcitedefaultendpunct}{\mcitedefaultseppunct}\relax
\EndOfBibitem
\bibitem[Landi \emph{et~al.}(2025)Landi, Russo, Sciortino, and
  Valeriani]{landi2024self}
C.~Landi, J.~Russo, F.~Sciortino and C.~Valeriani, \emph{Soft Matter}, 2025,
  \textbf{21}, 45--54\relax
\mciteBstWouldAddEndPuncttrue
\mciteSetBstMidEndSepPunct{\mcitedefaultmidpunct}
{\mcitedefaultendpunct}{\mcitedefaultseppunct}\relax
\EndOfBibitem
\bibitem[Noguchi and Gompper(2005)]{Noguchi2005}
H.~Noguchi and G.~Gompper, \emph{Phys. Rev. E}, 2005, \textbf{72}, 011901\relax
\mciteBstWouldAddEndPuncttrue
\mciteSetBstMidEndSepPunct{\mcitedefaultmidpunct}
{\mcitedefaultendpunct}{\mcitedefaultseppunct}\relax
\EndOfBibitem
\bibitem[Weeks \emph{et~al.}(1971)Weeks, Chandler, and Andersen]{weeks1971role}
J.~D. Weeks, D.~Chandler and H.~C. Andersen, \emph{The Journal of chemical
  physics}, 1971, \textbf{54}, 5237--5247\relax
\mciteBstWouldAddEndPuncttrue
\mciteSetBstMidEndSepPunct{\mcitedefaultmidpunct}
{\mcitedefaultendpunct}{\mcitedefaultseppunct}\relax
\EndOfBibitem
\bibitem[Jiang and Hou(2014)]{jiang2014motion}
H.~Jiang and Z.~Hou, \emph{Soft Matter}, 2014, \textbf{10}, 1012--1017\relax
\mciteBstWouldAddEndPuncttrue
\mciteSetBstMidEndSepPunct{\mcitedefaultmidpunct}
{\mcitedefaultendpunct}{\mcitedefaultseppunct}\relax
\EndOfBibitem
\bibitem[Foglino \emph{et~al.}(2019)Foglino, Locatelli, Brackley, Michieletto,
  Likos, and Marenduzzo]{Foglino2019}
M.~Foglino, E.~Locatelli, C.~A. Brackley, D.~Michieletto, C.~N. Likos and
  D.~Marenduzzo, \emph{Soft Matter}, 2019, \textbf{15}, 5995--6005\relax
\mciteBstWouldAddEndPuncttrue
\mciteSetBstMidEndSepPunct{\mcitedefaultmidpunct}
{\mcitedefaultendpunct}{\mcitedefaultseppunct}\relax
\EndOfBibitem
\bibitem[Janzen \emph{et~al.}(2025)Janzen, Miranda, Martín-Roca, Malgaretti,
  Locatelli, Valeriani, and Fernandez]{janzen2024active}
G.~Janzen, J.~P. Miranda, J.~Martín-Roca, P.~Malgaretti, E.~Locatelli,
  C.~Valeriani and D.~A.~M. Fernandez, \emph{The J. Chem. Phys}, 2025,
  \textbf{162}, 114905\relax
\mciteBstWouldAddEndPuncttrue
\mciteSetBstMidEndSepPunct{\mcitedefaultmidpunct}
{\mcitedefaultendpunct}{\mcitedefaultseppunct}\relax
\EndOfBibitem
\bibitem[Sknepnek(2024)]{SAMoS2024}
R.~Sknepnek, \emph{{SAMoS: Self-propelled Agent-based Models with SAMoS}},
  \url{https://github.com/sknepneklab/SAMoS}, 2024, Accessed: 2024-04-29\relax
\mciteBstWouldAddEndPuncttrue
\mciteSetBstMidEndSepPunct{\mcitedefaultmidpunct}
{\mcitedefaultendpunct}{\mcitedefaultseppunct}\relax
\EndOfBibitem
\bibitem[Leimkuhler and Matthews(2015)]{leimkuhler2015molecular}
B.~Leimkuhler and C.~Matthews, \emph{Molecular Dynamics: With Deterministic and
  Stochastic Numerical Methods}, Springer International Publishing, 2015\relax
\mciteBstWouldAddEndPuncttrue
\mciteSetBstMidEndSepPunct{\mcitedefaultmidpunct}
{\mcitedefaultendpunct}{\mcitedefaultseppunct}\relax
\EndOfBibitem
\bibitem[Krantz(1999)]{krantz1999handbook}
S.~G. Krantz, \emph{Handbook of Complex Variables}, Birkh{\"a}user Boston, MA,
  1st edn., 1999\relax
\mciteBstWouldAddEndPuncttrue
\mciteSetBstMidEndSepPunct{\mcitedefaultmidpunct}
{\mcitedefaultendpunct}{\mcitedefaultseppunct}\relax
\EndOfBibitem
\bibitem[SI()]{SI}
\emph{Supplementary Material.}\relax
\mciteBstWouldAddEndPunctfalse
\mciteSetBstMidEndSepPunct{\mcitedefaultmidpunct}
{}{\mcitedefaultseppunct}\relax
\EndOfBibitem
\bibitem[Mitchison and Kirschner(1984)]{mitchison1984}
T.~Mitchison and M.~Kirschner, \emph{Nature}, 1984, \textbf{312},
  237--242\relax
\mciteBstWouldAddEndPuncttrue
\mciteSetBstMidEndSepPunct{\mcitedefaultmidpunct}
{\mcitedefaultendpunct}{\mcitedefaultseppunct}\relax
\EndOfBibitem
\bibitem[Howard and Hyman(2003)]{howard2003}
J.~Howard and A.~A. Hyman, \emph{Nature}, 2003, \textbf{422}, 753--758\relax
\mciteBstWouldAddEndPuncttrue
\mciteSetBstMidEndSepPunct{\mcitedefaultmidpunct}
{\mcitedefaultendpunct}{\mcitedefaultseppunct}\relax
\EndOfBibitem
\bibitem[Wegner(1976)]{wegner1976}
A.~Wegner, \emph{Journal of Molecular Biology}, 1976, \textbf{108},
  139--150\relax
\mciteBstWouldAddEndPuncttrue
\mciteSetBstMidEndSepPunct{\mcitedefaultmidpunct}
{\mcitedefaultendpunct}{\mcitedefaultseppunct}\relax
\EndOfBibitem
\bibitem[Pollard and Borisy(2003)]{pollard2003}
T.~D. Pollard and G.~G. Borisy, \emph{Cell}, 2003, \textbf{112}, 453--465\relax
\mciteBstWouldAddEndPuncttrue
\mciteSetBstMidEndSepPunct{\mcitedefaultmidpunct}
{\mcitedefaultendpunct}{\mcitedefaultseppunct}\relax
\EndOfBibitem
\bibitem[Hansen and McDonald(1986)]{hansen1986theory}
J.-P. Hansen and I.~R. McDonald, \emph{Theory of Simple Liquids}, Academic
  Press, London, 2nd edn., 1986\relax
\mciteBstWouldAddEndPuncttrue
\mciteSetBstMidEndSepPunct{\mcitedefaultmidpunct}
{\mcitedefaultendpunct}{\mcitedefaultseppunct}\relax
\EndOfBibitem
\bibitem[Helfferich \emph{et~al.}(2018)Helfferich, Brisch, Meyer, Benzerara,
  Ziebert, Farago, and Baschnagel]{helfferich2018}
J.~Helfferich, J.~Brisch, H.~Meyer, O.~Benzerara, F.~Ziebert, J.~Farago and
  J.~Baschnagel, \emph{European Physical Journal E}, 2018, \textbf{41},
  71\relax
\mciteBstWouldAddEndPuncttrue
\mciteSetBstMidEndSepPunct{\mcitedefaultmidpunct}
{\mcitedefaultendpunct}{\mcitedefaultseppunct}\relax
\EndOfBibitem
\end{mcitethebibliography}
\bibliographystyle{rsc}

\end{document}